\documentclass[aps,prl,reprint,groupedaddress]{revtex4-1}

\usepackage{graphicx}
\usepackage{dcolumn}
\usepackage{rotating}
\usepackage{amssymb}
\usepackage{mathptmx}
\usepackage{amsfonts}
\usepackage{amsmath}
\usepackage{bm}
\begin{document}

. 
\title{Bose condensation and spin superfluidity of magnons in a perpendicularly magnetized film of yttrium iron garnet.}

\author{ P.\,M.\,Vetoshko$^{+*}$, G.\,A.\,Knyazev$^{+}$, A.\,N.\,Kuzmichev$^{+}$,  A.\,A.\,Holin$^{*}$,  
V.\,I.\,Belotelov$^{+*-}$,  
Yu.\,M.\,Bunkov$^{+}$\thanks{e-mail: y.bunkov@rqc.ru}\\~}

\affiliation{$^+$ Russian Quantum Center, Skolkovo, 143025, Moscow, Russia\\
	$^*$ Crimean Federal University. V.I. Vernadsky, 295007, Simferopol, Russia\\
$^-$Moscow State University M.V. Lomonosov, 119992, Moscow, Russia}

\begin{abstract}
The formation of a Bose condensate of magnons in a perpendicularly magnetized film of yttrium iron   garnet under radio-frequency pumping in a strip line is studied experimentally. The characteristics of nonlinear magnetic resonance and the spatial distribution of the Bose condensate of magnons in the magnetic field gradient are investigated. In these experiments, the Bosonic system of magnons behaves similarly to the Bose condensate of magnons in the antiferromagnetic superfluid $^3$He-B, which was studied in detail earlier. Magnonic BEC forms a coherently precessing state with the properties of magnonic superfluidity. Its stability is determined by the repulsive potential between excited magnons, which compensates for the inhomogeneity of the magnetic field. 
	 
\end{abstract}

\maketitle

Currently, macroscopic quantum phenomena are of great interest. It can be used to create platforms for quantum computing.
Recent success in Google's creation of a quantum computer based on superconducting qubits~\cite{QSC}
stimulated the search for other similar systems.
In particular, it is proposed to use the phenomenon of magnonic superfluidity as a basis for the magnon quantum processor~\cite{BunkovComp}.
In this article, we draw attention to a system of coherent magnons arising upon the excitation of nonlinear magnetic resonance in an iron-yttrium garnet (YIG) film magnetized perpendicular to the plane.
The dynamic properties of this system are in many aspects similar to the properties of magnons in superfluid $^3$He-B, in which magnonic superfluidity and coherent precession of magnetization  were discovered~\cite{LLIDS, Fomin}. In this article, we present to our readers an experiment on the formation of a state of coherently precessing magnons in YIG in a strongly inhomogeneous magnetic field, the results of which are analogous to the pioneering observations of this effect in $^3$He-B~\cite{Separ, Volovik20}.

Studies of YIG films under strong excitation conditions have led to the observation of the effect of nonlinear resonance, in which the precession frequency depends on the amplitude of its excitation (Foldover resonance)~\cite {Anderson}. An approximate analytical solution of the Landau-Lifshitz equations under nonlinear resonance conditions can be obtained only for the simplest case of a single oscillator in the limit of a relatively small frequency shift~\cite{TeorFoldover}. Real macroscopic samples have spatial inhomogeneity and should be described by a set of coupled oscillators.
The theoretical analysis is complicated due to the fact that the excitation of the resonance is also spatially inhomogeneous, especially when it is excited by a strip line. Also, in addition to the local Hilbert damping, one should also take into account the relaxation processes associated with spin diffusion in the case of spatial inhomogeneity of the resonance, as well as the interaction with the environment.
All these factors lead to the impossibility of constructing a theory that would actually describe the signals of nonlinear resonance in YIG~\cite{Fetisiv} films. The use of micromodeling programs is also limited, since the required simulation time must exceed the magnon lifetime, which for YIG can reach 100,000 precession periods. In addition, this time dramatically increases with an increase in the size of the sample. However, to describe the resonance in the case of a high excitation level, one can use the quantum properties of magnons, namely, the fact that magnons are Bose particles and condense into a magnonic Bose condensate (mBEC) at a sufficient concentration. Exactly that approach proposed in this article allows us to describe the basic properties of nonlinear resonance in magnetic materials.

Unlike atoms, the number of which is conserved, the density of magnons can change due to their creation and annihilation from the physical vacuum of a magnetically ordered state, in accordance with the Holstein-Primakov formalism~\cite{HP}.
At a low concentration, magnon gas can be considered as spin waves - an object of classical physics described by the Landau-Lifshitz equations.
At a finite temperature, the number of thermally activated magnons is determined by Bose statistics and is always below the critical formation density of mBEC. However, the magnon density can be significantly increased by exciting additional magnons from a magnetically ordered state (physical vacuum) using magnetic resonance methods. In this case, the deviation of the magnetization from the equilibrium direction corresponds to the production of magnons, the number of which is determined by the change in the longitudinal magnetization of the system, $N_r=(M_0-M_z)_r/\hbar $, where $N_r$ is the density of magnons, $(M_0-M_z)$ is the difference between the total magnetization and its projection onto the axis of the stationary magnetic field.
The magnon density required for the formation of a Bose condensate can be easily calculated for various magnetically ordered materials, as demonstrated in \cite{BunkovSafonov}. In particular, the critical magnon density in a perpendicularly magnetized YIG film corresponds to a deviation of the magnetization vector from normal by $2.5^\circ$, which under experimental conditions corresponds to an external field shift of about 2 Oe, thus, all of the described results, obtained at a pump power of more than 1 dBm, correspond to the conditions of the formed mBEC, according to statistics.

Strictly speaking, the properties of mBEC go beyond the scope of classical physics and are traditionally described by the Gross-Pitaevsky formalism, developed to describe the atomic Bose condensate \cite{Salazar}.
Magnonic BEK is a macroscopic quantum state described by a wave function: 
\begin{equation}
|\Psi|
_r={\mathcal N_r}^{1/2}  e^{i\mu
	t_r+i\alpha_r}\,,
\label{ODLRO}
\end{equation}
where $\mu$ and $\alpha$ are the chemical potential and the phase of the wave function, and $N_r$ is the density of excited magnons. The chemical potential of magnons is determined by their precession frequency and can be spatially inhomogeneous. The phase gradient of the wave function leads to a superfluid flux of magnons directed towards the region of a lower magnetic field, i.e., a lower chemical potential:
\begin{equation}
{\bf J} = {\mathcal N} \nabla\alpha.
\label{SpinCurrent1}
\end{equation}
In a perpendicularly magnetized YIG film, the precession frequency depends on the density of excited magnons ~\cite{Gulyaev2000}:
\begin{equation}
\omega_N  = \omega_0 - \gamma 4\pi M{_0}\cos{\beta},
\label{freqshiftNoRF}
\end{equation}
where $\omega_0-\gamma4\pi M {_0}$ is the precession frequency at low excitation, which is determined by the external field and the demagnetizing field.
We are considering here a sample of a sufficiently large size, where the magnitude of the exchange interaction cannot synchronize the precession frequency across the sample size.
The superfluid flux of magnons into the region with a lower effective magnetic field leads to an increase in their density and, accordingly, to an increase in the precession frequency.
This process continues until a homogeneous precession state is established.
That is, a state with coherent precession of magnetization, similar to the one that was discovered in$^3$ He-B~\cite{Neprerivka, Book}.
If the RF field excites magnons locally, as in the case of a strip line, then these magnons are transferred by a superfluid current to a region with a lower magnetic field until a precession frequency equal to the pumping frequency is established in it, as in experiments with superfluid $^3$He-B. This process is described in detail in~\cite{Formir}.
The formation of a state with coherent precession of magnetization radically simplifies the problem of theoretical description of nonlinear resonance.
In this case, we are dealing with a macroscopic magnon condensate filling the entire sample space, in which the effective field is less than the field corresponding to the pump frequency.
In this case, the magnetization precesses coherently at the pump frequency. This state does not depend on the pump power, which is a distinctive feature of mBEC formation. MBEC automatically adsorbs energy from the exciting field, corresponding to its losses~\cite{Neprerivka}, which depend on the magnitude of the  magnetization deviation.
The nonlinear resonance signal decays as soon as the pumping value cannot compensate for the relaxation.
In this article, we are the first to experimentally show the formation of such a state in a YIG film and, in particular, under conditions when a sufficiently large magnetic field gradient is applied.
\begin{figure}[htt]
	\includegraphics[width=0.45\textwidth]{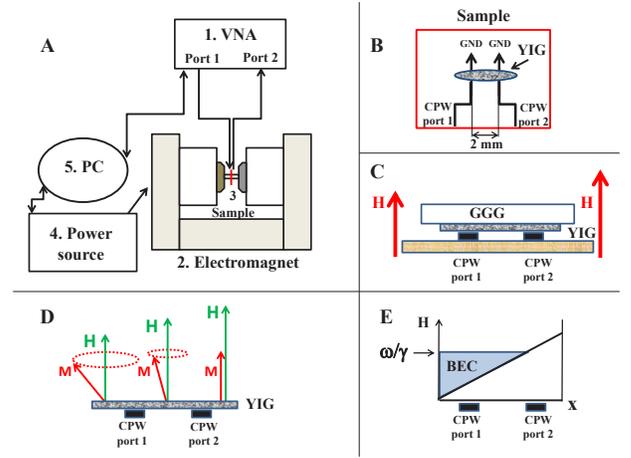}
	\caption{ Figure 1. Schematic of the setup (A), plateau with the sample, top view (B) and from the side (C), distribution of the angles of deflection of coherently precessing magnetization in the magnetic field gradient (D) and the area of filling the sample with magnonic BEC (E). }
	\label{shema}
\end{figure}
\begin{figure}[htt]
	\includegraphics[width=0.5\textwidth]{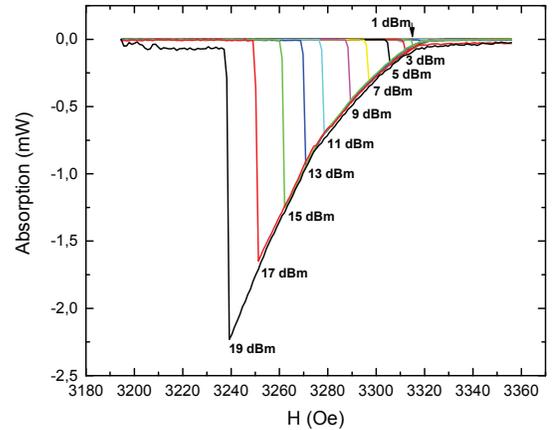}
	\caption{Figure 2. Change in the absorption of RF pumping in the first experiment with a decrease in the external magnetic field at different pumpimg powers indicated near the corresponding curves.}    
	\label{spectr}
\end{figure}
\begin{figure}[htt]
	\centering \includegraphics[width=0.45\textwidth]{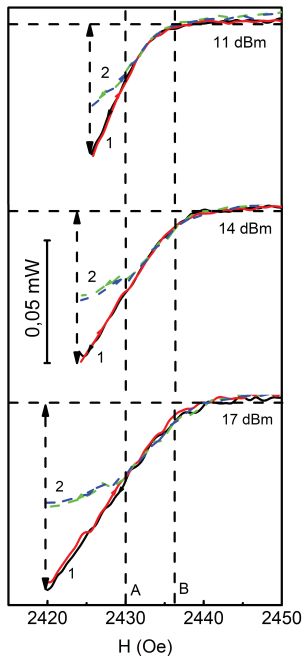}
	\caption{Figure 3. Change in the absorption of the RF power in the second experiment at different pumping powers. Lines 1 correspond to the signal from the first strip, and 2 - from the second. The amplitude of the signal from the second strip is increased by 5 times for comparison with the signal from the first.
		Signals are also shown when scanning the field upward. It is noteworthy that they coincide with the signals when scanning the field downward. Thus, in the case of a field gradient, hysteresis is not observed. }
	\label{spectr2}
\end{figure}

The general installation diagram is shown in Fig. 1 (A). The experiments were carried out on two samples of YIG films with a thickness of 6 $/mu$m and 0.6 $/mu$m. The first sample had the shape of a circle with a diameter of 0.5 mm, and the second, an ellipse with diameters of 5 and 0.5 mm. The films were grown at the Crimean Federal University. The first sample was located on one strip line, and the second - on two strip lines located symmetrically from the center of the sample at a distance of 2 mm from each other, as shown in Fig. 1 (B). The stripe line width was 0.2 mm. A magnetic field perpendicular to the film plane was applied to the sample. If in the first experiment the field was practically uniform, then in the second it had a gradient directed along the film, as shown in Fig. 1 (C). In the first case, the resonance was excited at a frequency of 3.7 GHz
and in the second - 1.856 GHz. In the experiment, we used the Keysite P5023A vector network analyzer. RF pumping was applied to the first strip.
Upon excitation of the resonance, the amplitude of the RF field in the strip decreased. Figure 2 shows the decrease in RF power when the field is scanned downward at different excitation powers.
An important observation is that the power absorbed by the sample is independent of the power applied to the strips.
This fundamentally contradicts the theory of nonlinear resonance~\cite{Anderson, TeorFoldover}, but fully corresponds to the properties of magnon BEC, previously investigated in $^3$He-B.
Indeed, the state of mBEC is completely determined by its chemical potential, which depends on the pump frequency, but not on its amplitude.
The pumping power determines the magnitude of the deviation of the precessing magnetization and the field shift at which its amplitude is not sufficient to compensate for the relaxation of magnons in mBEC.

Another property of mBEC is that it must fill the entire space in which the effective magnetic field is less than the corresponding pump frequency. To confirm this property of mBEC, we used a second sample and two stripe lines (Fig. 1 (B, C)).
The sample was placed in a magnetic field gradient of about 3.5 Oe per mm along the long axis of the sample (Fig. 1 (D, E)).
MBEC was excited by the first strip. In the case of a uniform field, the second strip received a signal from the first, which repeated the waveform on the first strip, reduced by approximately 5 times. It was not possible to distinguish this induced signal from the mBEC radiation signal. However, when the magnetic field gradient was applied, the signal from the second strip changed dramatically as shown in Fig. 3.
When the field is scanned downward, at point A, mBEC is formed in the region of the first strip. Upon further scanning of the field, the mBEC boundary, which corresponds to the condition $H=\omega/\gamma$ (see Fig. 1 (E)), moves to the second strip and reaches it at point B.
In this case, the second strip begins to receive the radiation signal from the mBEC. In Fig. 4 shows the power of the mBEC radiation signal received by the second rail. The region where the signal from the mBEC is received does not depend on the RF pump power, but only on the field difference in the region of the first and second stripes. This experiment directly demonstrates the spatial transport of magnons from the exciting region to the receiving strip.
This transfer is not possible in the case of a normal gas of magnons and is explained by the superfluid flux of magnons caused by the gradient of the chemical potential.
A similar experiment was carried out in superfluid $^3$He-B, in which the properties of the superfluid current of magnons between two RF coils~\cite{Ssupercurrent} were investigated.
It is also noteworthy that when the field is scanned backward, hysteresis does not appear, as shown in Fig. 3.
This effect of nonresonant excitation was previously noted in MnCO$_3$~\cite{Nonres}. The radio frequency field excites magnons in the modes of an inhomogeneous
resonance in the magnetic field gradient, which then condense into
homogeneous mBEC

\begin{figure}[htt]
	\includegraphics[width=0.45\textwidth]{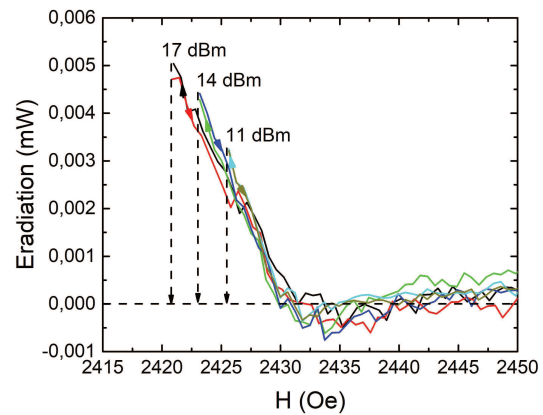}
	\caption{Figure 4. Signals of additional radiation from mBEC in the second strip as a function of the field at pumping powers of 17, 14 and 11 dBm. }
	\label{spectr4}
\end{figure}

The obtained results unambiguously confirm the formation of mBEC in the YIG film, magnetized perpendicularly. A signal of uniform precession on the second strip could arise only if the entire space with a lower field is filled with uniformly precessing mBEC. The results of studying YIG films in an EPR spectrometer also confirm the formation of mBEC ~\cite{YIG}.
Also, recently in YIG was discovered the formation of a long-lived induction signal~\cite{YIGPisma}, in many aspects similar to the signals in $^3$He-B~\cite{PIS, BunkovVolovik2007}. Finally, in several works with antiferromagnetic MnCO$_3$ and CsMnF$_3$ samples, several effects were also discovered that demonstrates the existence of mBEC and magnonic superfluidity~\cite{BECMagnon, BECMagnon1, BECMagnon2, Gold}.
Thus, we can conclude that magnonic superfluidity in solid-state magnets is of the same nature as in superfluid $^3$He-B, despite the fundamental difference in their ground state~\cite{SpinWaves, BEC2}.
 
It should be noted that the resulting state is superfluid in the sense that deviation from it causes the formation of precession phase gradients, which excite a superfluid magnon current, which flows until the gradients disappear and coherence is restored.
In a large series of experimental works with various phases of superfluid $^3$He, all magnonic analogs of known superfluid and superconducting effects, such as superfluid spin current in the channel and phase slip at its critical value~\cite{PhSlip, Ssupercurrent}, spin-current Josephson effect~\cite{Joseph, SPIN}, formation of quantum vortices at circular magnetization current~\cite{Vortex},
Goldstone vibration modes~\cite{Goldst, Goldst2, Goldst3}, etc.

At present, intensive studies are underway for another type of mBEC, which appears in the longitudinally magnetized YIG film~\cite {KS, KS2, KS3, KS4}. In this configuration, excited magnons attract. Therefore, homogeneous precession is unstable and decays into spin waves, as was shown experimentally for the superfluid $^3$He-A~\cite{Instab, Instab2}. Accordingly, there is no need to speak of a superfluid state, since the Landau critical velocity is zero. In the case of longitudinal magnetization of the YIG film, the minimum energy corresponds to running magnons, which form the nontrivial mBEC. If the mBEC considered in the first case corresponds to the classical atomic BEC of resting atoms, then for the second type of mBEC there is no analogy in the world of particles. However, it also remains a very interesting subject for research and applications.
 
This work was supported by the Ministry of Science and Higher Education of the Russian Federation, Megagrant No. 075-15-2019-1934.

\end{document}